\documentclass[twocolumn,showkeys,prd,nofootinbib,floatfix,preprintnumbers]{revtex4-1}

\usepackage[utf8]{inputenc}
\usepackage{amsfonts,amsmath,amssymb} 
\usepackage{graphicx,graphics,color}
\usepackage{gensymb}
\usepackage{longtable}
\usepackage{bbding}

\begin{document}

\preprint{PI/UAN-2020-668FT}

\title{Coupled Multi-Proca Vector Dark Energy}

\author{L.~Gabriel G\'omez$^{1}$ and Yeinzon Rodr\'iguez$^{1,2,3}$}

\affiliation{$^1$Escuela de F\'isica, Universidad Industrial de Santander, Ciudad Universitaria, Bucaramanga 680002, Colombia}
\affiliation{$^2$Centro de Investigaciones en Ciencias B\'asicas y Aplicadas, Universidad Antonio Nari\~no,
Cra 3 Este \# 47A-15, Bogot\'a D.C. 110231, Colombia}
\affiliation{$^3$Simons Associate at The Abdus Salam International Centre for Theoretical Physics,
Strada Costiera 11, I-34151, Trieste, Italy}


\begin{abstract}
We study a new class of vector dark energy models where multi-Proca fields $A_\mu^a$ are coupled to cold dark matter by the term $f(X)\tilde{\mathcal{L}}_{m}$ where $f(X)$ is a general function of $X\equiv -\frac{1}{2}A^\mu_ a A^a_\mu$ and $\tilde{\mathcal{L}}_{m}$ is the cold dark matter Lagrangian. From here, we derive the general covariant form of the novel interaction term sourcing the field equations. This result is quite general in the sense that encompasses Abelian and non-Abelian vector fields. In particular, we investigate the effects of this type of coupling in a simple dark energy model based on three copies of canonical Maxwell fields  to realize isotropic expansion. The cosmological background dynamics of the model is examined by means of a dynamical system analysis to determine the stability of the emergent cosmological solutions. As an interesting result, we find that the coupling function leads to the existence of a novel scaling solution during the dark matter dominance. Furthermore, the critical points show an early contribution of the vector field in the form of dark radiation and a stable de Sitter-type attractor at late times mimicking dark energy. The cosmological evolution of the system as well as the aforementioned features are  verified by numerical computations. Observational constraints are also discussed to put the model in a more phenomenological context in the light of future observations.

\end{abstract}

\pacs{Valid PACS appear here}
\maketitle


\section{Introduction}

Our current understanding of the universe is based on the standard model of cosmology, in short $\Lambda$CDM ($\Lambda$ Cold Dark Matter) \cite{Aghanim:2018eyx}. Under this realization, the universe is experiencing today an accelerated expansion due to a constant energy density characterized by some repulsive pressure and attributed to the cosmological constant \cite{Riess:1998cb,Schmidt:1998ys,Perlmutter:1998np}. Despite its success, the model has been the subject of intense debate due to some controversies and problems when confronted with observations, such as the cosmological constant problem \cite{Weinberg:1988cp,Amendola:2015ksp} and the inference of some spatial curvature suggested by the lensing
amplitude in the cosmic microwave background power spectra \cite{DiValentino:2019qzk}. Likewise, recent observations of the redshift-space distorsion \cite{Macaulay:2013swa} and cluster counts \cite{Battye:2014qga,Alam:2016hwk,Abbott:2017wau} have pointed out a lower rate for the cosmic growth than that predicted by the $\Lambda$CDM model. 

However, perhaps one of the major concerns in the scientific community about the standard model of cosmology, among the aforementioned internal inconsistencies, is the lack of conciliation between early and late time universe measurements which has been referred to as the  Hubble tension \cite{Freedman:2017yms}. The Planck experiment infers a value  $H_{0}= 67.4\pm 0.5$~ km/s/Mpc at $68\%$ C.L., assuming the standard $\Lambda$CDM model \cite{Aghanim:2018eyx} while the recent value determined from the observation of long-period Cepheids in the Large Magellanic Cloud \cite{Riess:2019cxk} is $H_{0}= 74.03\pm1.42$~km/s/Mpc at $68\%$ C.L., thereby putting the standard model in tension.

Three major routes have been established, among the numerous theoretical proposals (see e.g refs. \cite{Amendola:2015ksp,Clifton:2011jh,Ezquiaga:2018btd} and references therein), to reconcile such discrepancies and other issues in the $\Lambda$CDM model. These are, modifying the geometric sector of Einstein gravity by breaking its fundamental assumptions \cite{Lovelock:1971yv,Horava:2009uw,deRham:2014zqa,Heisenberg:2018vsk}, including extra fields minimally coupled to gravity which leads to dynamical dark energy (DE) such as quintessence \cite{Wetterich:1987fm,Ratra:1987rm} and k-essence \cite{ArmendarizPicon:1999rj,ArmendarizPicon:2000dh,ArmendarizPicon:2000ah}, or extending gravity by building non-minimal interactions between matter and gravity\footnote{It is important to clarify that Horndeski theories include all the generalized Jordan-Brans-Dicke type varieties, such as quintessence and k-essence \cite{Ezquiaga:2018btd}.} \cite{Heisenberg:2018vsk,Horndeski:1974wa,Horndeski:1976gi,Nicolis:2008in,Deffayet:2011gz,Kobayashi:2011nu}. Yet, within the second possibility, interactions between dark mater and DE are allowed. This idea has been explored intensively through phenomenological interactions introduced into the conservation equations. We are not going to discuss, however, the vast amount of possible choices studied in the literature; instead, we refer to refs. \cite{Bahamonde:2017ize,Wang:2016lxa} and references therein for the cosmological implications of non-trivial functional forms of couplings. It has been shown, however, that this artificial description may introduce both inconsistencies with the covariant stress energy conservation and instabilities  \cite{Tamanini:2015iia}. Hence, one must appeal for a more fundamental and robust approach to account for such a coupling within the dark sector based, for instance, on a field theoretical description. Under this perspective, several consistent approaches have been developed to build coupled dark energy models at the level of the action such as conformal/disformal transformations \cite{Bekenstein:1992pj,Amendola:1999qq,Gleyzes:2015pma}, scalar-fluid  theories \cite{Boehmer:2015kta,Boehmer:2015sha} and the post-Friedmannian  formalism \cite{Skordis:2015yra}. Yet another possibility is to couple directly the DE field to the matter Lagrangian, i.e., at the Lagrangian level, through a coupling function as has been proposed in a pioneering work called coupled quintessence \cite{Amendola:1999er}. This type of coupling, and more general non-trivial ones, can arise naturally after an appropriate conformal transformation which relates the metric in the Jordan frame,  where matter lives, with the Einstein frame, where gravity is described by general relativity  \cite{Amendola:1999qq}. No matter the underlying physical origin of the coupling, that is,  whether it comes from high energy physics arguments or from another yet unknown physical reason, one can motivate the interaction with the aim of addressing some inconsistencies in the standard cosmological model  with the phenomenological interest of testing their observational signatures. 

It is important to mention that most of the existing (coupled) dark energy models are based on scalar fields with little or no presence of space-like vector fields.  This is because, unlike time-like vector fields, space-like vector fields can generate large amounts of anisotropy which is inconsistent with observations \cite{Aghanim:2018eyx}. Nevertheless, one may avoid this undesirable result, for instance, by taking a large number of random vector fields so that, on average, they lead to an isotropic universe \cite{Golovnev:2008cf} or by considering a set of three space-like vector fields of the same norm and pointing towards mutually orthogonal spatial directions, shaping what is called the cosmic triad \cite{ArmendarizPicon:2004pm}. We highlight, up to the best of our knowledge, some significant progress in the construction of coupled vector dark energy models based on space-like vector fields such as the cosmic triad cosmology \cite{Zhao:2008tk}, three-form dark energy models \cite{Koivisto:2012xm,Ngampitipan:2011se,Yao:2017enb}, extensions of the pioneering vector-like dark energy model \cite{ArmendarizPicon:2004pm} through phenomenological interactions \cite{Wei:2006tn,Landim:2016dxh} and a direct  coupling driving anisotropic expansion \cite{Koivisto:2008xf}. It is worth mentioning that a coupled vector dark energy model has been proposed recently \cite{Nakamura:2019phn} in the context of the generalized Proca theory \cite{Tasinato:2014eka,Heisenberg:2014rta,Allys:2015sht,Allys:2016jaq,Jimenez:2016isa} with a trivial time-like vector configuration to comply with the background symmetry. 

We study in this work a novel coupling between multiple vector fields and matter of the form $f(X)\tilde{\mathcal{L}}_{m}$, where $f(X)$ is a general function of $X\equiv -\frac{1}{2}A^\mu_ a A^a_\mu$ and $\tilde{\mathcal{L}}_{m}$ is the matter Lagrangian, following closely the spirit of the kinetically coupled scalar dark energy scenario \cite{Barros:2019rdv}. Our main purpose is to extend the kinetic scalar coupling to an analogous vector coupling  which is fully independent of the assumed model and on the Abelian/non-Abelian nature of the vector fields. To
investigate its cosmological  implications, we frame this coupling into a model consisting of three copies of the standard Maxwell-theory, consistent with the isotropic background, with a specific power law form for the coupling and an exponential potential. 

The content of this paper is structured as follows. We derive, in section \ref{sec:2}, the general interaction term for the proposed coupling function. 
In section \ref{sec:3}, we study its cosmological behaviour by means of a dynamical system analysis in a specific setup consisting of the cosmic triad compatible with a Friedmann-Lema$\hat{\rm{\i}}$tre-Robertson-Walker (FLRW) universe. We confirm the obtained qualitative features by numerically solving the cosmological evolution of the system in section \ref{sec:4}. Finally, the discussion and perspectives of this work are presented in section \ref{sec:5}.

\section{The model}\label{sec:2}

We start by writing the general $\mathcal{L}_{2}$ piece of the Lagrangian for the generalized multi-Proca theory \cite{Allys:2016kbq,Jimenez:2016upj,GallegoCadavid:2020dho} that involves only gauge-invariant and spontaneous symmetry breaking quantities which are present in the Abelian and non-Abelian cases. Indeed, any function of a set of vector fields $A_{\mu}^a$, their associated gauge field strength tensors $F_{\mu\nu}^a$, and their Hodge duals $\tilde{F}_{\mu\nu}^a$ belongs to $\mathcal{L}_2$\footnote{Here $a$ labels the different vector fields and becomes a group index if there exists some internal symmetry group.}:
\begin{equation}
    \mathcal{L}_{2}\equiv f(A_{\mu}^{a},F_{\mu\nu}^{a},\tilde{F}_{\mu\nu}^{a}).\label{sec:2:eqn1}
\end{equation}
Thus, the action involving the Einstein-Hilbert Lagrangian $\mathcal{L}_{EH}$, the $\mathcal{L}_{2}$ piece, and the explicit coupling between a mass-type term $X \equiv -\frac{1}{2}A_\mu^a A^\mu_a$ of the vector fields and dark matter reads
\begin{equation}
   \mathcal{S}=\int d^{4}x\;\sqrt{-g}(\mathcal{L}_{EH}+\mathcal{L}_{2}+f(X)\mathcal{\tilde{L}}_{m}),\label{sec:2:eqn2.0}
\end{equation}
where $g$ is the determinant of the metric.
The novel contribution of this paper is the inclusion of the coupling function\footnote{A similar coupling was studied in the context of the generalized Proca theory with a time-like vector field \cite{Nakamura:2019phn}. However, our approach differs from it in that we derive
here the associated
interacting terms.} $f(X)$ that describes the way in which multi-Proca fields $A_\mu^a$ couple to dark matter.  The latter, in turn, is described by the Lagrangian $\tilde{\mathcal{L}}_{m}(g_{\mu\nu},\psi)$, where $\psi$ is the dark matter field. 
%
In particular, we consider the action
%
\begin{equation}
   \mathcal{S}=\int d^{4}x\;\sqrt{-g} \left(\frac{M_{p}^2}{2}R -\frac{1}{4}F_{\mu\nu}^{a}F_{a}^{\mu\nu}-V(\tilde{X})+ f(X)\mathcal{\tilde{L}}_{m}\right),\label{sec:2:eqn2}
\end{equation}
where $M_p$ is the reduced Planck mass, $R$ is the Ricci scalar, $F_{\mu\nu}^{a} \equiv \nabla_{\mu}A_{\nu}^{a}-\nabla_{\nu}A_{\nu}^{a}$, and $V$ is a potential term, as in the uncoupled model of ref. \cite{ArmendarizPicon:2004pm}, where we have defined the quantity $\tilde{X}\equiv A_{\mu}^{a} A^{\mu}_{a}$.

We assume for simplicity that radiation and baryons are completely uncoupled, otherwise they both would feel a metric different to the one that is identified with the gravitational sector. In such a case, one would move from one representation to the other by means of a conformal/disformal transformation \cite{Bekenstein:1992pj}. 

On the other hand, it is interesting to note that, in contrast to phenomenological interactions, the coupling at the level of the action, which can be identified as a purely vector conformal transformation \cite{Kimura:2016rzw}, affects not only the continuity equations of each matter component but also the gravitational field equations\footnote{Note, however, that the coupling  naturally becomes
ineffective in regions of low dark matter density and vanishes
in the absence of dark matter.}. Hence, by varying the action in eqn.~(\ref{sec:2:eqn2}) with respect to the metric, we obtain the gravitational field equations
\begin{equation}
    \frac{M_{p}^{2}}{2}G_{\mu\nu}=T_{\mu\nu}^{A}+f \tilde{T}_{\mu\nu}^m+f_{,X}\mathcal{\tilde{L}}_{m} A_{\mu}^{a}A_{\nu a},\label{sec:2:eqn3}
\end{equation}
where the energy-momentum tensor of the multi-Proca fields is 
\begin{align}
T_{\mu\nu}^{A}&\equiv -2\frac{\delta\mathcal{L}_{A}}{\delta g^{\mu\nu}}+\mathcal{L}_{A} g_{\mu\nu} \nonumber\\
    &=F_{\mu\sigma}^{a}F_{a\nu}^{\sigma}+2 V_{,\tilde{X}}A_{\mu}^{a}A_{\nu a}-\left(\frac{1}{4}F_{\rho\sigma}^{a}F^{\rho\sigma}_{a}+V(\tilde{X})\right)g_{\mu\nu},\label{sec:2:eqn4.0}
\end{align}
$\mathcal{L}_A$ being defined by 
\begin{equation}
    \mathcal{L}_A \equiv -\frac{1}{4}F_{\mu\nu}^{a}F_{a}^{\mu\nu}-V(\tilde{X}).
\end{equation}
Note that we have introduced for abbreviation the definitions $f_{,X}\equiv \frac{df}{dX}$ and $V_{,\tilde{X}}\equiv \frac{dV}{d\tilde{X}}$ in the above expressions. It is convenient to redefine the dark matter energy-momentum tensor according to the gravitational field equations in eqn.~(\ref{sec:2:eqn3}) such that
\begin{align}
     T_{\mu\nu}^{m}\equiv &-2\frac{\delta\mathcal{L}_{m}}{\delta g^{\mu\nu}}+\mathcal{L}_{m} g_{\mu\nu} \nonumber \\
    =&   f \tilde{T}_{\mu\nu}^m+ f_{,X} \mathcal{\tilde{L}}_{m} A_{\mu}^{a} A_{\nu a},\label{sec:2:eqn5}
\end{align}
where we have defined for convenience $\mathcal{L}_{m}\equiv  f\mathcal{\tilde{L}}_{m}$ and introduced the definition
\begin{equation}
    \tilde{T}_{\mu\nu}^{m}\equiv -2\frac{\delta\tilde{\mathcal{L}}_{m}}{\delta g^{\mu\nu}}+\tilde{\mathcal{L}}_{m} g_{\mu\nu} .\label{sec:2:eqn5.1}
 \end{equation}
Now, the field equations become
\begin{equation}
    \frac{M_{p}^{2}}{2}G_{\mu\nu}=T_{\mu\nu}^{A}+T_{\mu\nu}^{m}.
\end{equation}
Note that we have passed over the arguments in all the functions for brevity. It is worth pointing out that the new dark matter energy-momentum tensor $T_{\mu\nu}^{m}$ can have, apparently, an additional non-vanishing pressure component if the vector fields have spatial components by virtue of the second term in eqn.~(\ref{sec:2:eqn5}). On the contrary, if we considered a time-like vector, there would be an additional contribution to the energy density as usually happens in kinetically coupled scalar models (see e.g. ref. \cite{Barros:2019rdv}). Hence, $T_{\mu\nu}^{m}$ seems to be ambiguously defined 
in the sense that
it depends on the components of the different $A_{\mu}^{a}$. We shall discuss later this point and its implications to have a right interpretation of the cold dark matter component.

The coupled $A_\mu$ field equations of motion are obtained after varying the action with respect to $A^a_\mu$. These can be written in the reduced form
\begin{equation}
    \partial_{\mu}(\sqrt{-g}F^{a\mu\nu})=2\sqrt{-g}A^{\nu a} \left( V_{,\tilde{X}}+\frac{1}{2}\frac{f_{,X}}{f}\mathcal{L}_{m}\right).\label{sec:2:eqn7}
\end{equation}

As the conservation law sets the interacting term sourcing the continuity equation for each component without ambiguity, we find from here the general interacting expression for the coupling function in eqn.~(\ref{sec:2:eqn2}) by taking the covariant derivative of eqn.~(\ref{sec:2:eqn5}):
\begin{align}
    \nabla_{\mu}T^{\mu m}_{\nu}=&-A_{\beta}^{c}\nabla_{\mu}A^{\beta}_{c}\Big[\frac{f_{,X}}{f}(T^{\mu m}_{\nu}-f_{,X}A^{\mu a}A_{\nu a}\mathcal{L}_{m})\nonumber \\
    &+\frac{f_{,XX}}{f}\mathcal{L}_{m}A^{\mu}_{a}A_{\nu}^{a}\Big]+
\frac{f_{,X}}{f}A^{\mu}_{c}A_{\nu}^{c}\Big[\nabla_{\mu}\mathcal{L}_{m}\nonumber
 \\ &+\frac{f_{,X}}{f}\mathcal{L}_{m} A_{\beta}^{a}\nabla_{\mu}A^{\beta}_{a}\Big]+\frac{f_{,X}}{f}\mathcal{L}_{m}[A_{\nu}^{c}\nabla_{\mu}A^{\mu}_{c}\nonumber \\ &+A^{\mu}_{c}\nabla_{\mu}A_{\nu}^{c}],\label{sec:2:eqn8}
\end{align}
such that $\nabla_{\mu}T^{\mu m}_{\nu}= -\nabla_{\mu}T^{\mu A}_{\nu}$. Note that we have guaranteed the conservation of the uncoupled dark matter fluid $\nabla_{\mu}\tilde{T}^{\mu m}_{\nu}=0$. It is worth noting that the general form of the interaction term depends explicitly on the vector fields and their first-order derivatives. In addition, it is valid for Abelian and non-Abelian vector fields since the coupling function only depends on $A^a_{\mu}$ and not, for instance, on the field strength tensor which involves the specific structure of the associated group. This alternative, however, suggests that we can construct more general coupling functions, for instance, a coupling of the form $f(Y)$ where
\begin{equation}
    Y \equiv -\frac{1}{4}F_{\mu\nu}^{a}F_{a}^{\mu\nu},
\end{equation}
so that the structure of the associated group becomes explicit\footnote{For the particular case of SU(N) gauge fields, the self-interaction term that contributes to the kinetic energy, it being proportional to the coupling constant of the  group, will appear now explicitly in both the gravitational and the $A_\mu$ field equations.}. This possibility clearly deserves further examination as well as the possibility of settling this idea in the context of vector-tensor theories through conformal/disformal transformations as manifestation of possible non-minimal couplings to gravity when moving to the Jordan frame. Interesting frameworks to have in mind were recently developed for  metric transformations based on a U(1) gauge field \cite{Papadopoulos:2017xxx}, on the field strength tensor \cite{Gumrukcuoglu:2019ebp}, and on its dual tensor \cite{DeFelice:2019hxb}.

\subsection{Particular setup: The cosmic triad}
We consider a triad of mutually orthogonal and of the same norm space-like vector fields $A^{\mu}_{a}$ which is 
an arrangement compatible with isotropic and homogeneous background:
\begin{equation}
    A_{\mu}^{a}\equiv a(t) A(t) \delta_{\mu}^{a}.\label{sec:3:eqn1}
\end{equation}
Here $A(t)$ is the common norm of the three vector fields and $a(t)$ is the scale factor in the FLRW spacetime with line element $ds^{2}=-dt^{2}+a^{2}(t)\delta_{ij}dx^{i}dx^{j}$. Bearing this in mind, it is possible to compute the gravitational field equations and the $A_\mu$ field equations where the presence of the coupling function $f$ is revealed:\footnote{This realization results in an interesting generalization of the pioneering model of ref. \cite{ArmendarizPicon:2004pm} and opens up a wide possibility to build coupled dark energy models driven only by vector fields.}
\begin{equation}
3M_{p}^{2}H^{2}=f  \tilde{\rho}_{m}+\frac{3}{2}(\dot{A}+HA)^{2}+V,\label{sec:3:eqn2}
\end{equation}
\begin{equation}
M_{p}^{2}(3H^{2}+2\dot{H})=-\frac{1}{2}(\dot{A}+HA)^{2}+V-2 V_{,\tilde{X}}A^{2}+f_{,X}A^{2}\tilde{\rho}_{m},\label{sec:3:eqn3}
\end{equation}
\begin{equation}
\ddot{A}+\left( \frac{\ddot{a}}{a}+H^{2}\right)A+3H \dot{A}+2V_{,\tilde{X}}A-f_{,X}A \tilde{\rho}_{m}=0.\label{sec:3:eqn4}
\end{equation}
Here an upper dot denotes a derivative with respect to the cosmic time, $H(t)\equiv \dot{a}/a$ is the Hubble parameter, and $\tilde{\rho}_m$ is the cold dark matter energy density. As we assume a  (pressureless) cold dark matter fluid, the last term in eqn.~(\ref{sec:3:eqn3}) can be interpreted as an effective pressure for the vector field  arising from the coupling to matter. Consequently, we define the density and  pressure of the vector field respectively as
\begin{equation}
\rho_{A}=\frac{3}{2}(\dot{A}+HA)^{2}+V,\label{sec:3:eqn5}
\end{equation}
\begin{equation}
p_{A}=\frac{1}{2}(\dot{A}+HA)^{2}-V+2V_{,\tilde{X}}A^{2}-\frac{f_{,X}}{f}A^{2}\rho_{m},\label{sec:3:eqn6}
\end{equation}
where we have introduced the definition $\rho_{m}\equiv f\tilde{\rho}_{m}$. It will be instructive to keep in mind that the energy density of the vector field is the result of the contribution of the kinetic term $Y$ and the vector field potential $V$ as in the canonical case;  however, the pressure has an additional contribution due to the coupling function. Such a contribution will be important whenever $\rho_m$ does not vanish, affecting consequently the total energy budget of the universe (see eqn.~\ref{sec:3:eqn2}). Indeed, as we shall see, the kinetic term will contribute early as dark radiation $p_{A}=p_{Y}=\frac{1}{3}\rho_{Y}$ as expected commonly for space-like vector fields, whereas the potential will work as DE to account for the late-time accelerated expansion.

From eqns.~(\ref{sec:3:eqn5})-(\ref{sec:3:eqn6}),  it is possible to write the $A_\mu$ field equation (eqn.~\ref{sec:3:eqn4}) in the form of the continuity equation for a fluid, allowing us to infer the interacting term:
\begin{equation}
\dot{\rho}_{A}+3H(\rho_{A}+p_{A})=3A\dot{A}\frac{f_{,X}}{f}\rho_{m},\label{sec:3:eqn7}
\end{equation}
\begin{equation}
\dot{\rho}_{m}+3H\rho_{m}=-3A\dot{A}\frac{f_{,X}}{f}\rho_{m}.\label{sec:3:eqn7.2}
\end{equation}
The continuity equations evidence clearly that  both components interact with each other through a novel interaction term $Q=-\frac{3f_{,X}}{f}\rho_{m}$ which is very similar to the one derived from the coupled tachyonic dark energy model \cite{Teixeira:2019tfi}.

As we are mostly interested in the late-time cosmology, we define the effective state parameter from  eqn.~(\ref{sec:3:eqn3}) as follows:

\begin{equation}
w_{\rm eff}\equiv \frac{p_{\rm T}}{\rho_{\rm T}}=-\left(1+\frac{2\dot{H}}{3H^{2}
}\right),\label{sec:3:eqn8}
\end{equation}
where $p_T$ and $\rho_T$ are respectively the total pressure and energy density,
and search accordingly for the condition $w_{\rm eff}<-1/3$ in such a period. This parameter will describe the transition from the domination of each component in accordance to the evolution of the universe. It is also important to define the state parameter for the vector field as $w_{A}\equiv \frac{p_{A}}{\rho_{A}}$ which will coincide with $w_{\rm eff}$ at early and late times as we will see. We have thereby provided all the key equations needed to describe the cosmological background dynamics.  We will first address this issue in a qualitative way by using the dynamical systems approach in the next section.  Then, we will numerically solve the coupled system to visualize the cosmic evolution in section \ref{sec:4}.

\begin{figure}[!htb]
\centering
\includegraphics[width=0.8
\hsize,clip]{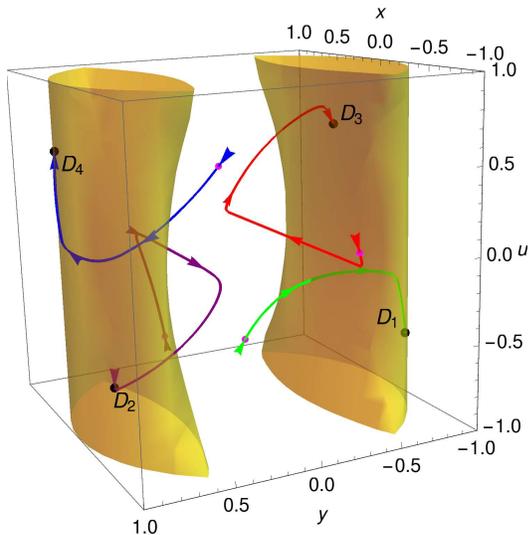}
\caption{Trajectories for different initial conditions (cyan points) in phase space converging to the attractor solutions $D_{1,2,3,4}$ (black points). These trajectories correspond to numerical solutions of the dynamical system in eqn.~(\ref{sec:4:eqn3}), after dimensional reduction, for the parameters $q=-0.1$ and $\lambda=0.4$. The light yellow region denotes the physical region we define as the intersection of the physical phase space and the region where the universe undergoes a standard
accelerated expansion ($w_{eff}< -1/3$), the latter's being more restrictive.}\label{fig:1}
\end{figure}
\begin{table*}[htp]
\centering
\caption{Fixed points of the autonomous system in eqn.~(\ref{sec:4:eqn3}) and their main physical features. The dynamical stability is found by demanding the standard requirement of negativity of the real parts of the eigenvalues of the Jacobian matrix $\mathcal{M}$ associated to the linear system evaluated at the critical points (see Appendix). In addition, the density parameter associated to the cosmic triad, the latter's state parameter, the effective state parameter, the conditions for the existence of the critical points in phase space, and the conditions for supporting late-time accelerated expansion are displayed as indicated.}
\begin{ruledtabular}
\begin{tabular}{cccccccccc}
Point & $x_{c}$ & $y_{c}$ & $z_{c}$ & $u_{c}$ & $\Omega_{A}$ & $w_{A}$ & $w_{\rm eff}$ & \text{Existence} & \text{Acceleration}
  \\ \hline
 $A_{\pm}$&$\pm 1$&$0$ &$0$&$\pm \sqrt{2}
$ & $1$ &$1/3$ &$1/3$ & $\forall q,\lambda$ & $\text{No}$\\
 $B_{\pm}$&$\pm\frac{\sqrt{2q}}{\sqrt{-1+2q}}$
 &$0$&$\pm\frac{1}{\sqrt{1-2q}}$&$\pm\sqrt{2}x_{c}$ & $\frac{2q}{-1+2q}$ & $0$ & $0$ & $q\neq1/2,\forall\lambda$ & $\text{No}$\\
 $D_{1,2}$&$-\frac{\sqrt{-1+3 \lambda}}{\sqrt{3\lambda}}$&$\mp\frac{1}{\sqrt{3\lambda}}$
 &$0$&$-\sqrt{2}x_{c}$ & $1$ & $-1$ & $-1$ & $\forall q, \lambda>1/3$ & $\forall q, \lambda$\\
$D_{3,4}$ &$\frac{\sqrt{-1+3 \lambda}}{\sqrt{3\lambda}}$&$\mp\frac{1}{\sqrt{3\lambda}}$
 &$0$&$\sqrt{2}x
_{c}$ & $1$ & $-1$ & $-1$ & $\forall q, \lambda>1/3$ & $\forall q, \lambda$\\
\end{tabular}
\end{ruledtabular}\label{tab:table1}
\end{table*}
%

\section{Dynamical system}\label{sec:3}
In order to study the background evolution of the model given by eqns.~(\ref{sec:3:eqn2})-(\ref{sec:3:eqn4}) and eqns.~(\ref{sec:3:eqn7})-(\ref{sec:3:eqn7.2})  by means of the dynamical system analysis, we transform the system of equations into an autonomous system of first-order differential equations.  To do so, we introduce the following dimensionless variables that will define the phase space  portrait:
\begin{align}
& x \equiv \sqrt{\frac{ (\dot{A}+HA)^{2}}{2M_{p}^{2}H^{2}}}; \;\; y\equiv \sqrt{\frac{V}{3M_{p}^{2}H^{2}}}; \nonumber \\
& z \equiv \sqrt{\frac{\rho_{m}}{3M_{p}^{2}H^{2}}}; \;\; u\equiv \frac{A}{M_{p}}.
\label{sec:4:eqn1}
\end{align}
These variables satisfy, in turn, the Friedmann constraint 
\begin{equation}
x^{2}+y^{2}+z^{2}=1.
\end{equation}
At this point, it is necessary to define the functional forms of both the potential and the coupling function.  For the former, we set an exponential potential while, for the latter, we set a power law\footnote{The reason for this choice is merely technical as it makes it easier to write the autonomous system in a closed way, rendering, in turn, the physical space compact. This fact guarantees that trajectories do not extend to infinity.}:
\begin{equation}
V(\tilde{X})=V_{0}e^{-\lambda \tilde{X}/M_{p}^{2}},\; f(X)=\left(\frac{X}{M_{p}^{2}}\right)^{q},
\label{sec:4:eqn2}
\end{equation}
such that $\frac{V_{,\tilde{X}}}{V}$ and $\frac{f_{,X}}{f}$ are completely determined as demanded to close the system of coupled equations. Here $\lambda$ and $q$ are the dimensionless model parameters to be constrained. With all this at disposition,  we derive the autonomous system 
\begin{align}
x^\prime&=-x\left(2+\frac{H^\prime}{H} \right) \nonumber \\
&+\frac{1}{x}\left[3\lambda y ^{2}u^{2
}-\sqrt{2}\frac{x}{u}q z^{2}
-y\left(y^\prime+y\frac{H^\prime}{H}\right)\right],\nonumber \\
y^\prime&=-y\left[\frac{H^\prime}{H}+3u^{2}\lambda\left(\sqrt{2}\frac{x}{u}-1\right) \right],\nonumber \\
z^\prime&=-z\left[\frac{H^\prime}{H}+\frac{3}{2}-q\left(\sqrt{2}\frac{x}{u}-1\right) \right],\nonumber \\
u^\prime&=\sqrt{2}x-u. \label{sec:4:eqn3}
\end{align}
Here the prime denotes a derivative with respect to $N\equiv\ln a$. We also determine the accelerating equation eqn.~(\ref{sec:3:eqn3}) in terms of the dimensionless quantities:
\begin{equation}
\frac{H^\prime}{H}=-\frac{3}{2}(1+w_{eff}),\label{sec:4:eqn4}
\end{equation}
with
\begin{equation}
w_{eff}=\frac{x^{2}}{3}-y^{2}-2\lambda y^{2}u^{2}+\frac{2}{3}q z^{2},\label{sec:4:eqn5}
\end{equation}
where the presence of the interacting term via the parameter $q$ and the quantity $z$ in the above equation is clear, evidencing thus the coupling of the vector field to matter. The state parameter for the vector DE is, in the same fashion,
\begin{equation}
w_{A}=\frac{x^{2}/3-y^{2}-2\lambda y^{2}u^{2}+\frac{2}{3}q z^{2}}{x^{2}+y^{2}}.\label{sec:4:eqn6}
\end{equation}
To ease our computations, we reduce the  4-dimensional phase space to a 3-dimensional one through the Friedmann constraint. The  physical volume of the reduced phase space becomes then  $x^{2}+y^{2}\leq1$ (with $u$ unconstrained). In what follows, we will compute the critical points of the system and study the general conditions that determine both their dynamical character and their existence in phase space, following the standard procedure of the dynamical systems theory.

\subsection{Phase space trajectories}
In order to better comprehend the dynamical character of the appearing solutions, we have first drawn some parametric trajectories in phase space for different initial conditions (marked as cyan points) in fig.~\ref{fig:1}. These points are initially close to a radiation dominated universe. As time passes the trajectories bring near some points 
(presumably saddle in nature)
before being finally
attracted towards four possible attractor solutions given by the critical points with coordinates $(x_{c,i},y_{c,i},u_{c,i})$ \footnote{Indeed, these solutions are doubly degenerate in the sense that they all represent one single physical solution of cosmological interest due to the symmetry of the invariant phase space. The emergence of these solutions, and their differences, are due to different sign combinations for the fixed points $(x_{c,i},y_{c,i},u_{c,i})$ with $i$ running from 1 up to 4, as can be seen in Table.~\ref{tab:table1}.}. The latter, as seen in fig.~\ref{fig:1}, are located in accordance to the odd-parity invariance of the system (see Table~\ref{tab:table1}) and coloured black inside the physical region. Such a physical region is denoted by light yellow and it is defined as the intersection space of the physical reduced phase space $x^2+y^2\leq 1$ and the region where the universe undergoes a standard
accelerated expansion ($w_{eff} < -1/3$), with $u$ now constrained by eqn.~(\ref{sec:4:eqn5}). One couple of solutions are, as a first criterion, located according to the sign of the parameter $u$. For instance $D_{1,2}$ correspond to $u$ negative while, for the opposite sign,  we have the attractors $D_{3,4}$. 
Nevertheless, they all have the same physical meaning: de Sitter-type attractors characterized by the dominance of vector DE through its potential. We have chosen $q=-0.1$ and $\lambda=0.4$ for these numerical solutions.  However, we have checked that the location of the attractor solutions inside the physical region does not depend on $q$ but does on $\lambda$.   Indeed, by increasing $\lambda$ significantly, the attractor points are slightly shifted from $y=\pm 1$ towards $y=0$, these being now enclosed by the renewed cylinder-shaped regions. Regarding the effect of $q$, it turns out to be important only in an intermediate stage of the universe when matter starts dominating.  This leads to a non-vanishing vector energy density contribution that can alter the dynamic expansion other than having some impact on non-linear processes of the early universe as the formation of cosmic structures. All of the above dynamical features will be more evident in the two-dimensional stream plots that we will display next as well as in the performed numerical computation that will be described in section \ref{sec:4}.
\begin{figure}[!htb]
\centering
\includegraphics[width=0.8\hsize,clip]{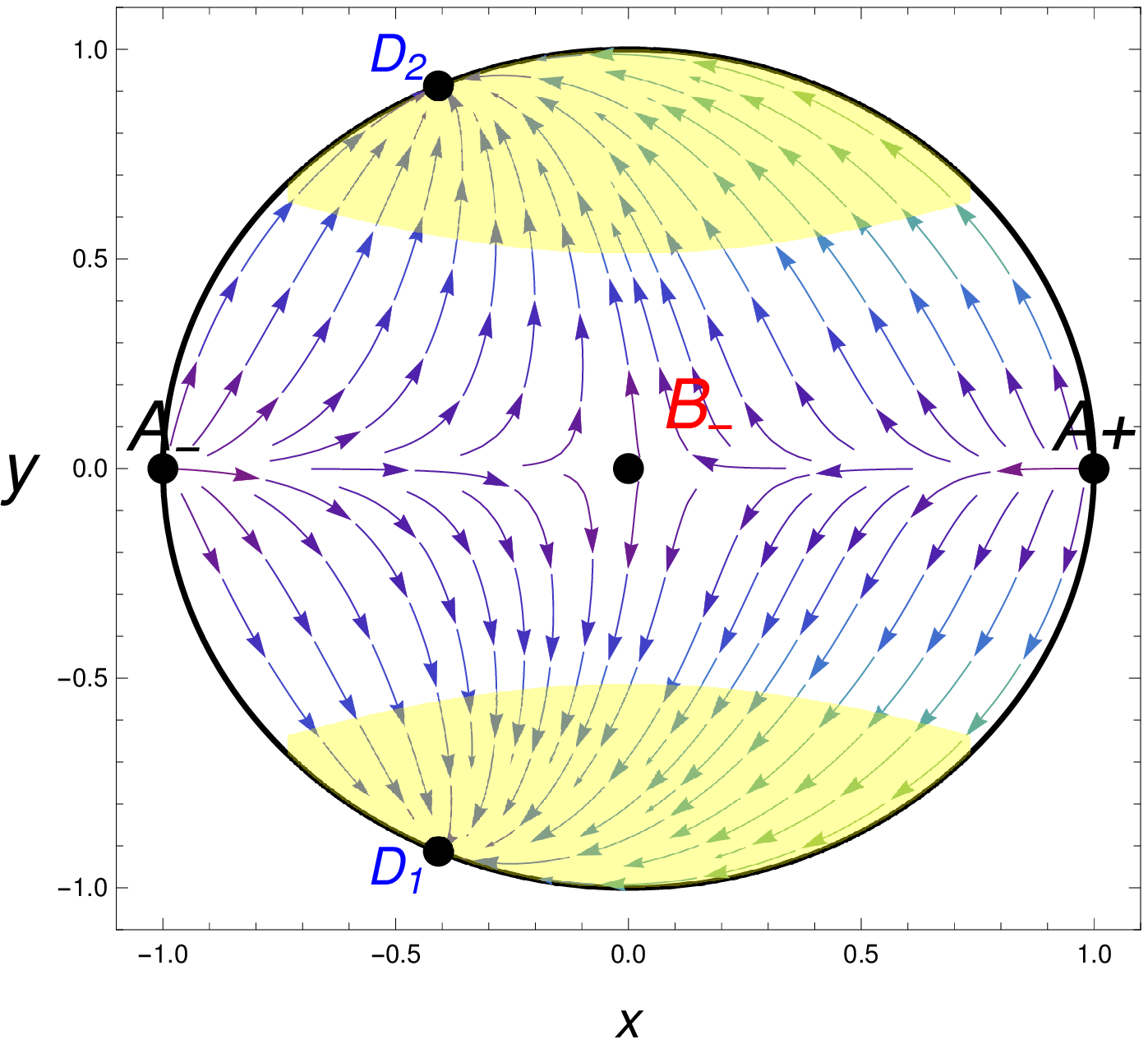}
\includegraphics[width=0.8\hsize,clip]{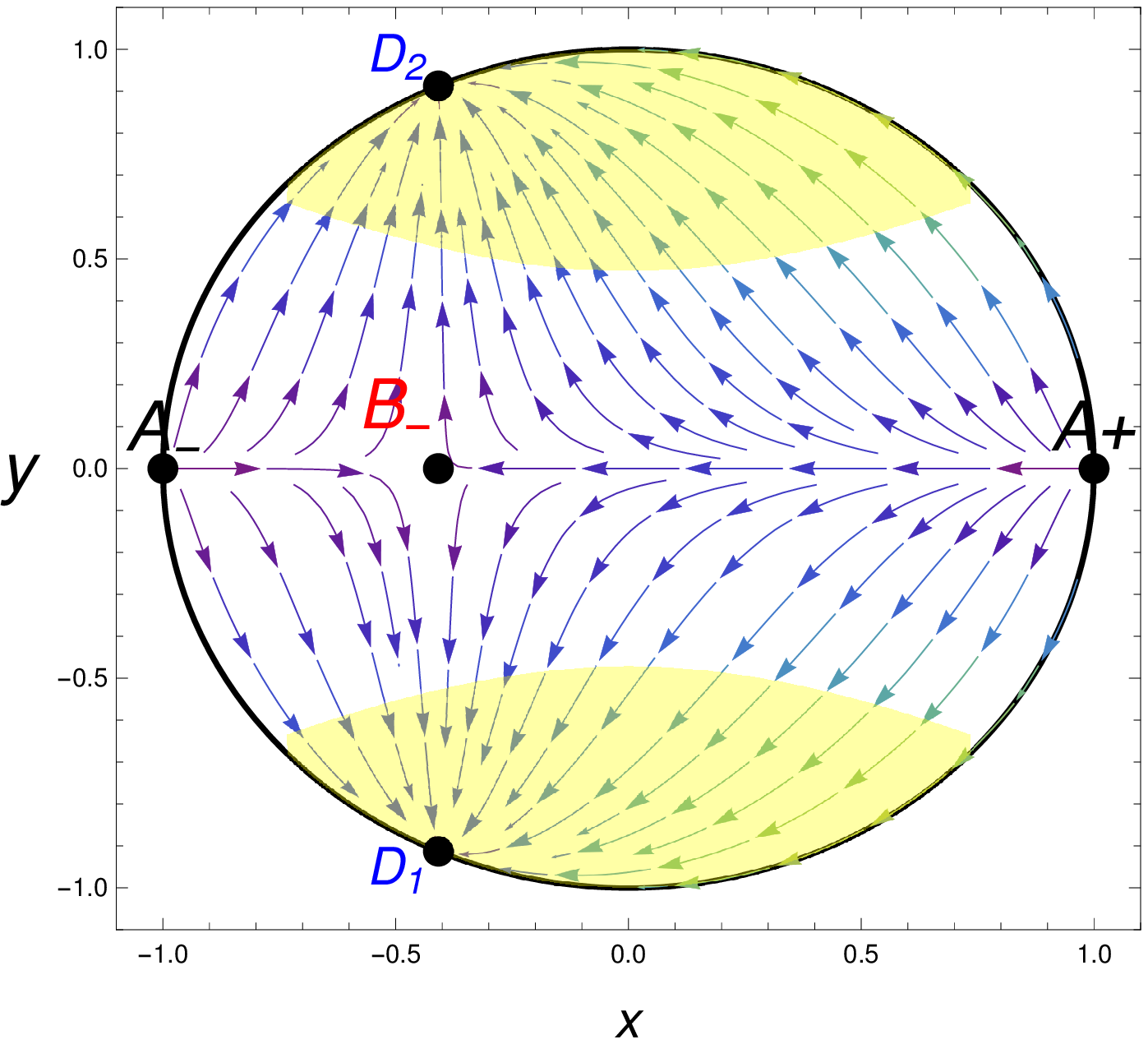}
\includegraphics[width=0.8\hsize,clip]{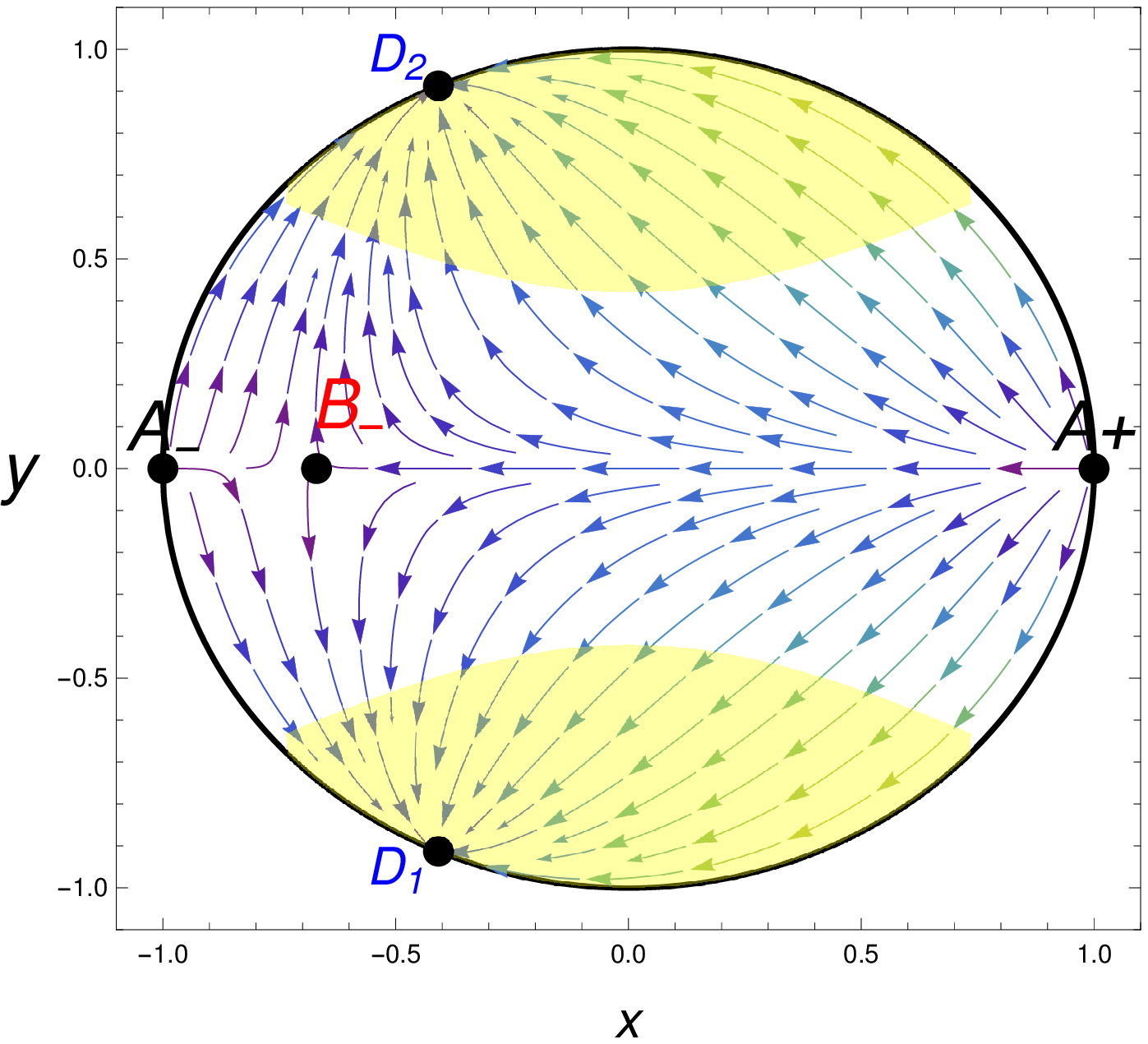}
\caption{Reduced phase space of the dynamical system. We have set the slice $u_{c}=-|u_{c}|=-\frac{\sqrt{-2/3+2\lambda}}{\sqrt{\lambda}}$, corresponding to $D_{1,2}$ and have taken $\lambda=0.4$ and $q=0,-0,1,-0.2$ in the top, middle and bottom panels respectively. The light yellow region denotes, as before, the physical region where the condition $x^2+y^2\leq 1$ is satisfied along with the condition for the accelerated expansion.}\label{fig:2}
\end{figure}
\begin{figure*}
\includegraphics[width=0.4\hsize,clip]{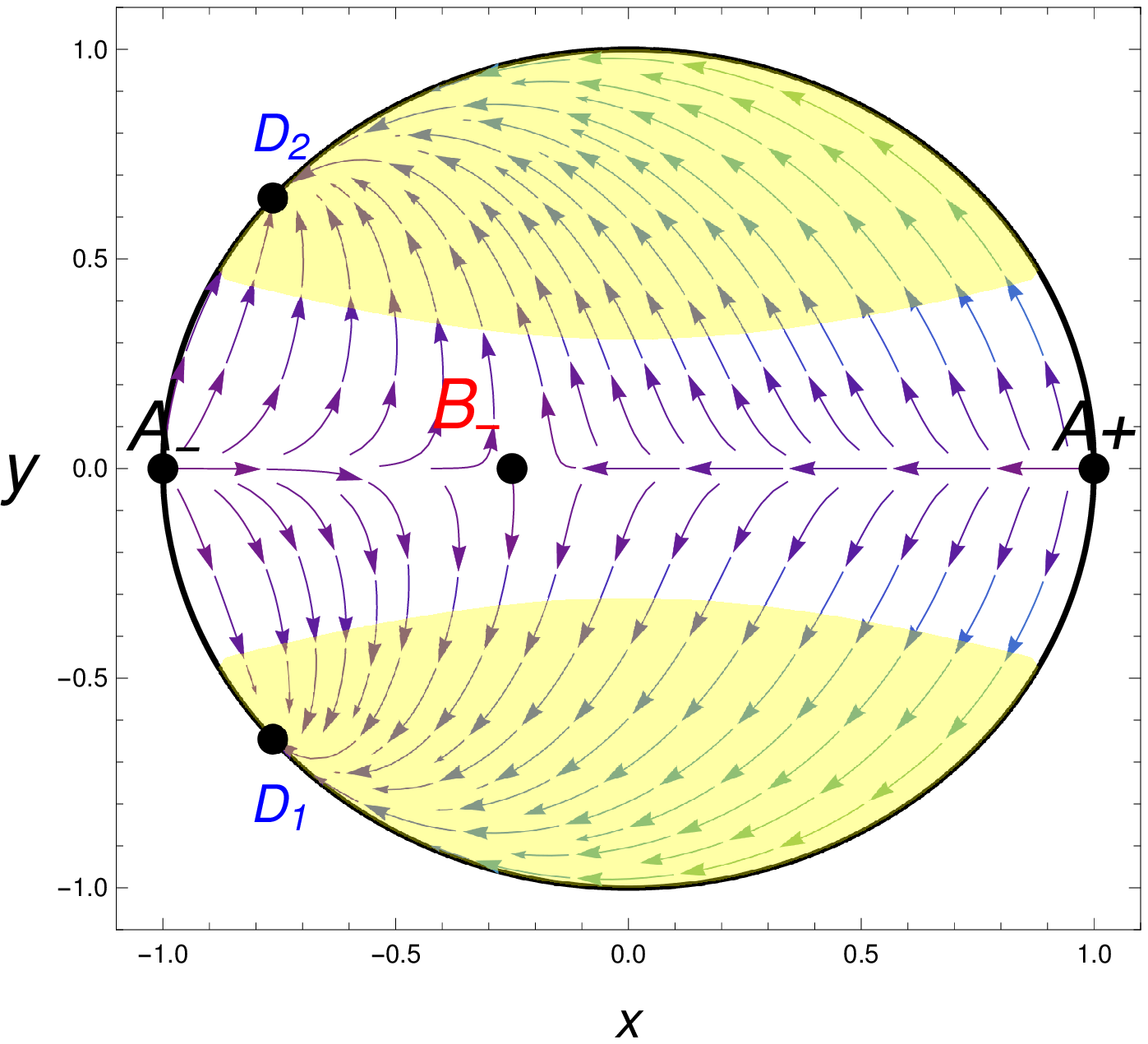}
\includegraphics[width=0.4\hsize,clip]{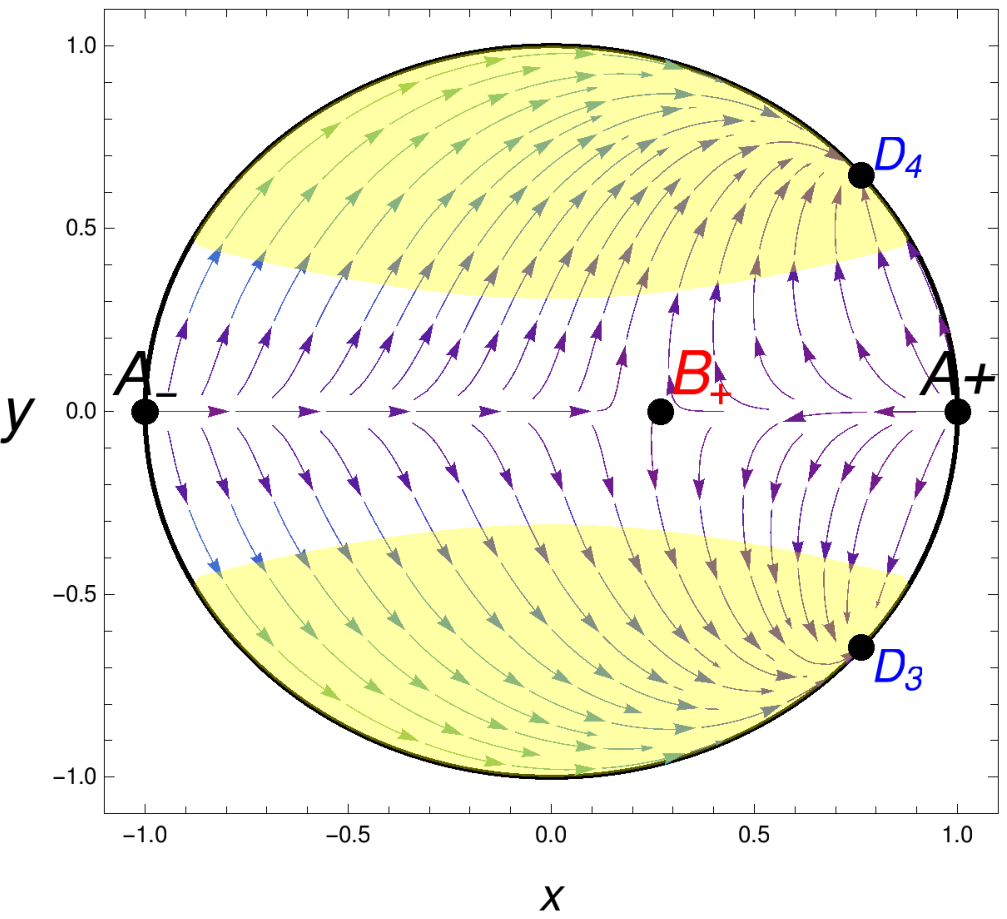}
\caption{Reduced phase space of the dynamical system. We have set the slice $u_{c}=-|u_{c}|$ for the left panel and $u_{c}=|u_{c}|$ for the right panel in order to  appreciate the equivalence between these two solutions which correspond to the attractors $D_{1,2}$ and $D_{3,4}$, respectively. It is also important to note how the attractor points approach  $A_{\pm}$ when we choose a larger value for $\lambda$ ($\lambda=0.8$) while keeping $q$ fixed  ($q=-0.1$). The light yellow region denotes the physical region where the universe is accelerating.}\label{fig:3}
\end{figure*}

\subsection{Critical points and stability}\label{sec:3.2}
The critical points are computed by matching to zero each equation of the autonomous system and solving a set of simple algebraic expressions. The solutions are shown in Table~\ref{tab:table1} corresponding to eight critical points. However, as we have discussed previously, they can be reduced without ambiguities to three physical solutions of cosmological interest, each one playing a crucial role in the evolution of the universe. We will refer then to a single solution when they exhibit the same physical features. Some physical quantities are also displayed in the table for a better description of the dynamical behaviour of the system such as the vector density parameter, the vector state parameter, the effective state parameter, the conditions for the existence of the critical points, and whether they correspond to an accelerated expansion period. The existence is 
determined by the requirement that $0\leq x^{2}+y^{2}\leq1$
and that the coordinates of the critical points must be real valued. Interestingly, there exists a new scaling solution mediated by the parameter $q$ (fixed point $B_{\pm}$). The emergence of this solution is due to the presence of the kinetic energy $x_{c}$ of the vector field during the matter dominated epoch contrary to the uncoupled case $q=0$ ($x_{c}=0$). This remarkable difference may lead to interesting implications not only at the background level, aiming at addressing the coincidence problem and the Hubble tension, but also at the linear/non-linear regime in perturbations, affecting for instance the formation and evolution of cosmic structures in the universe. This point will be discussed later  in view of the cosmological observations. We discuss the main physical features of each fixed point as follows: 

\begin{itemize}
\item \textit{Kinetic dominated solution} ($A_{\pm}$): this \textit{saddle point} corresponds to a kinetic dominated solution or dark radiation since $x$ is the only non-zero dynamical variable of the system. Hence, the cosmic triad fully dominates the energy density budget in the form of  dark radiation with $w_{A}=1/3$  and $\Omega_{A}=1$. We conclude that this point represents a \textit{radiation dominated period} with effective state parameter $w_{\rm eff}=1/3$ and that acceleration is not realizable during this initial stage. As can be seen in Table~\ref{tab:table1} and the Appendix, there is no explicit dependence of the existence and the stability conditions of this critical point on the parameter values.  

\item \textit{Kinetic scaling solution} ($B_{\pm}$): Though this point exists in the uncoupled case ($q=0$), the total energy density of the universe is not totally governed by matter ($\Omega_{m}=\frac{1}{1-2q}$) in this intermediate stage when $q \neq 0$. Instead, we have an outstanding contribution of the vector field ($\Omega_{A}=\frac{2q}{-1+2q}$ with $q\neq 1/2$) due to its coupling to matter through the parameter $q$. We may refer to this fixed point as the \textit{coupled multi-Proca vector-matter point}. Despite the contribution of the vector field to the total energy budget, this \textit{saddle point} represents  matter domination with $w_{eff}=0$ and with the vector field acting as a dust-like fluid.  
This shows that acceleration is not realizable unless the potential ($y_{c}$) dominated solution is achieved. On the other hand, this point exists for $q\neq1/2$ and for all values of $\lambda$. In addition, this preliminary inspection tells us that, in order to have well defined (real valued) positive energy density values for both components, the parameter $q$ must be negative. Nonetheless, feasible numerical solutions include also positive $q$ values as we will see. The most noticeable physical aspect of this solution is that it represents a novel scaling solution where the dark energy density to matter energy density ratio scales as $\rho_{DE}/\rho_{m}\propto -2q$, and has interesting implications at the background level aiming to solve the coincidence problem. This solution may be seen as a generalization of the pioneering  dark energy model of ref. \cite{ArmendarizPicon:2004pm} which is based on the cosmic triad.

\item \textit{Vector dominated solution} ($D_{i}$): This  critical point is a \textit{dynamical stable attractor} with all negative eigenvalues.  Interestingly, this point exhibits an accelerated period with state parameter $w_{A}=w_{eff}=-1$ independent entirely of the model parameters. Despite the explicit dependence of $w_{eff}$ on $q$ and $\lambda$ in eqn.~(\ref{sec:4:eqn5}), there appears an intriguing cancellation when replacing this critical point. This renders, as an attractor solution, a de Sitter universe with $w_{eff}=-1$ at late times as shall be verified numerically. In other words, the vector field mimics DE at late times with density parameter $\Omega_{A}=x_{c,i}^{2}+y_{c,i}^{2}=1$ being also independent of the model parameters.
\end{itemize}

In summary, the model we propose in this work provides a cosmological evolution where the cosmic triad contributes in the form of dark radiation at early times (kinetic dominated solution),  with a significant presence of its kinetic energy  quantified by the coupling $q$ at the matter domination era (kinetic scaling solution), and mimicking DE at late times (low redshift) through the exponential potential (vector dominated solution).

\begin{figure}[!htb]
\centering
\includegraphics[width=1\hsize,clip]{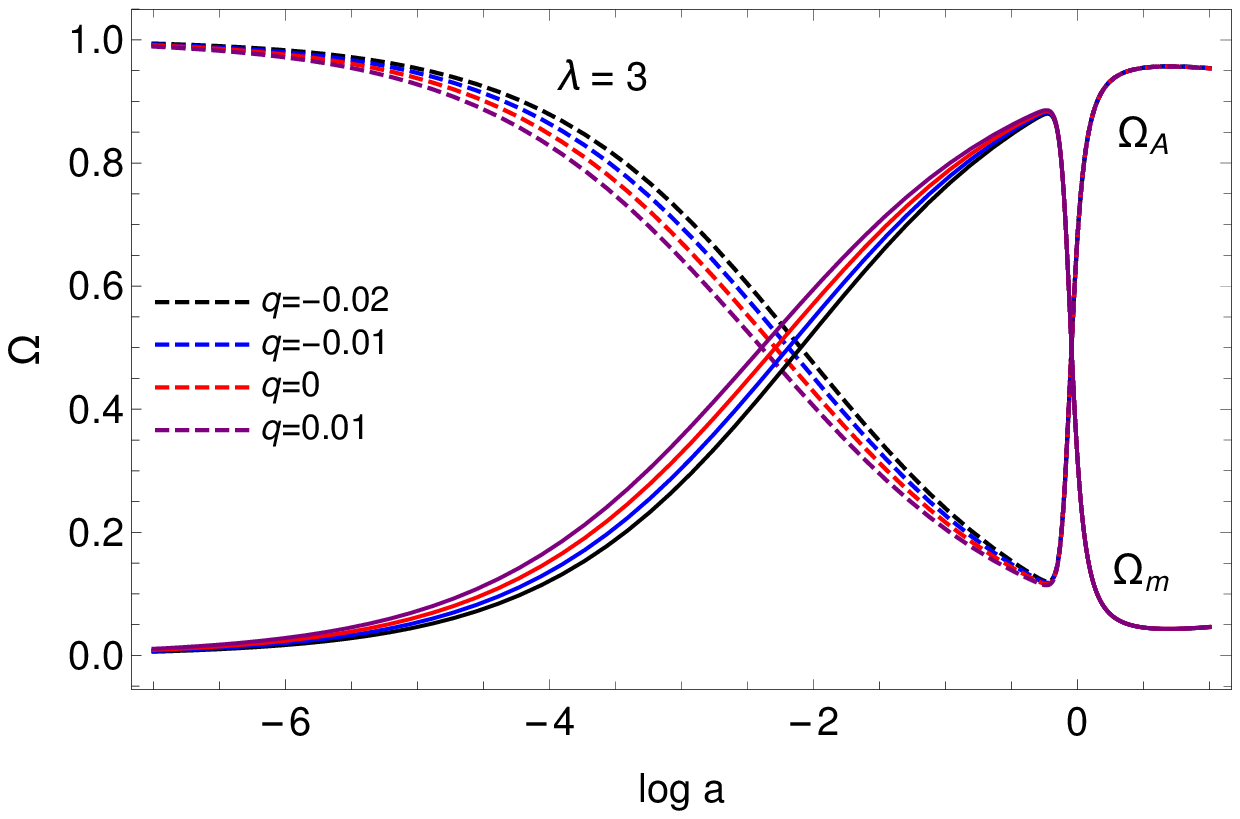}
\includegraphics[width=1\hsize,clip]{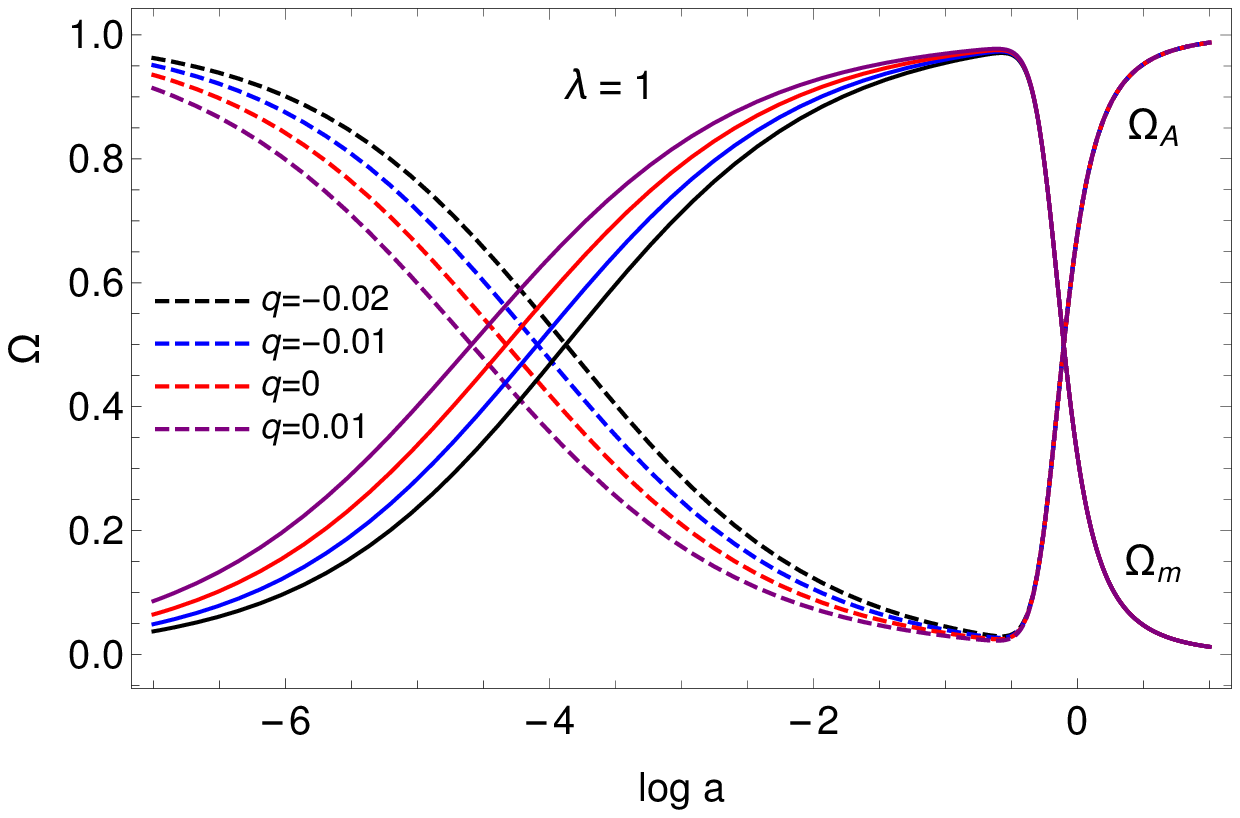}
\includegraphics[width=1\hsize,clip]{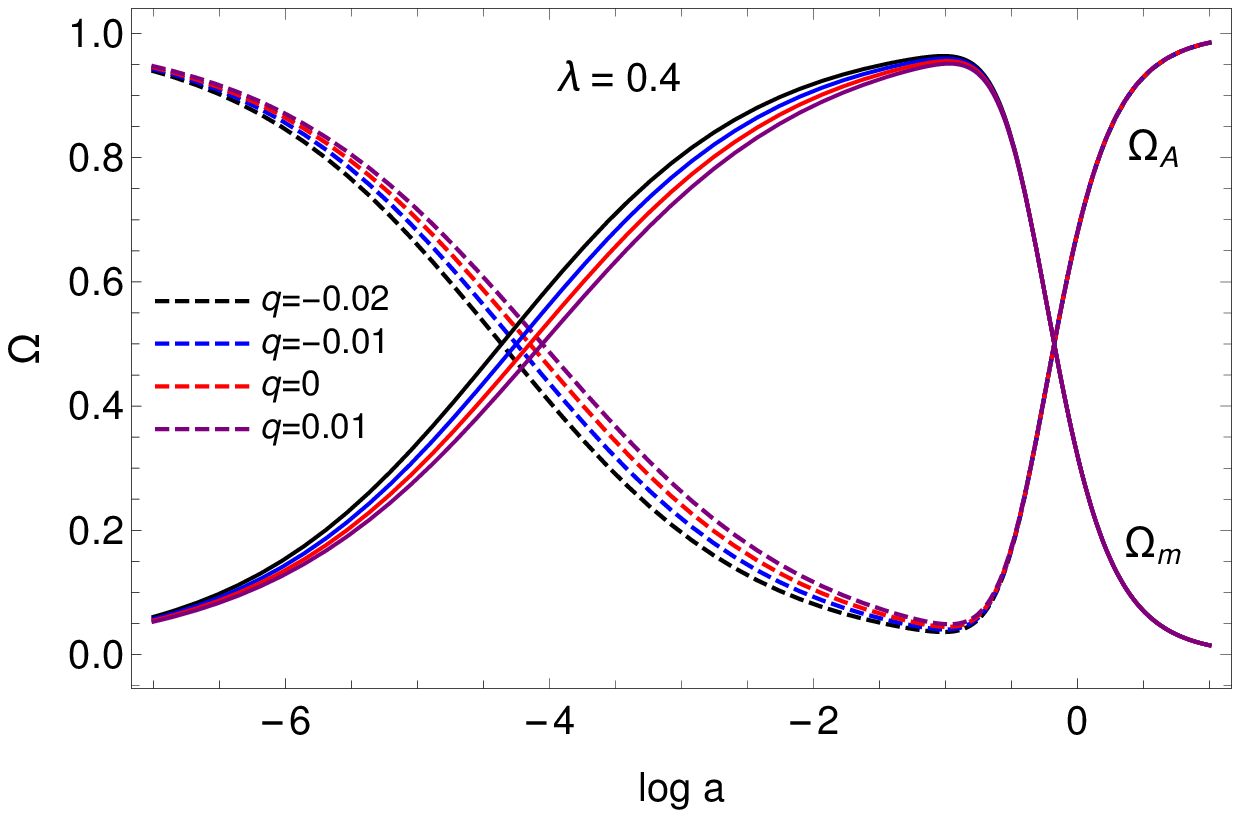}
\caption{Evolution of 
$\Omega_{A}$ and 
$\Omega_{m}$ versus the number of e-folds $N \equiv \log{a}$ for different values of $q$. 
Dashed and solid lines denote solutions for $\Omega_{A}$ and $\Omega_{m}$ respectively. Top, middle and bottom panels correspond to $\lambda=3,1$, and 0.4 respectively. We have chosen $u^{(0)}=0.9$ as a comparable value with respect to the $x^{(0)}$ and $y^{(0)}$ values.  The other initial conditions have been chosen to match the present value $\Omega_{A}=\Omega_{DE}^{(0)}=0.68$.} \label{fig:4}
\end{figure}
\begin{figure}
\centering
\includegraphics[width=1\hsize,clip]{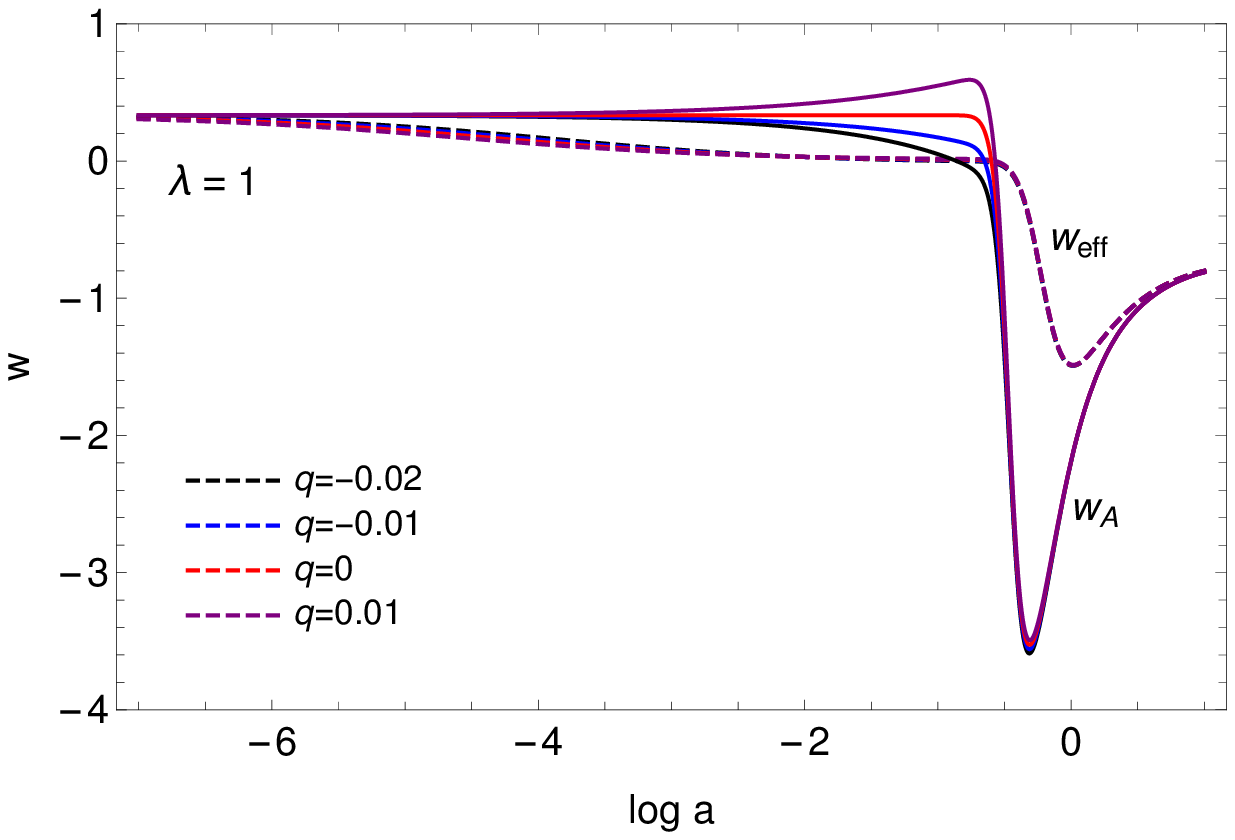}
\includegraphics[width=1\hsize,clip]{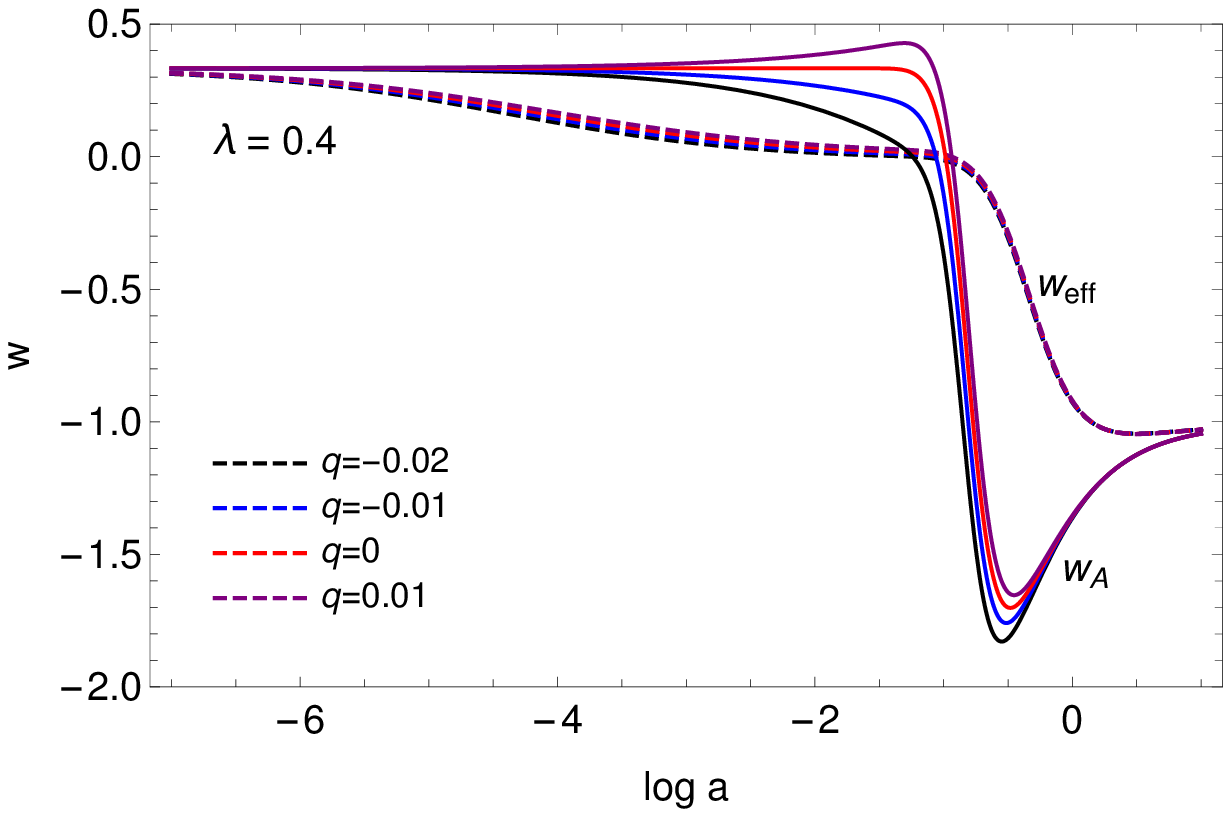}
\includegraphics[width=1\hsize,clip]{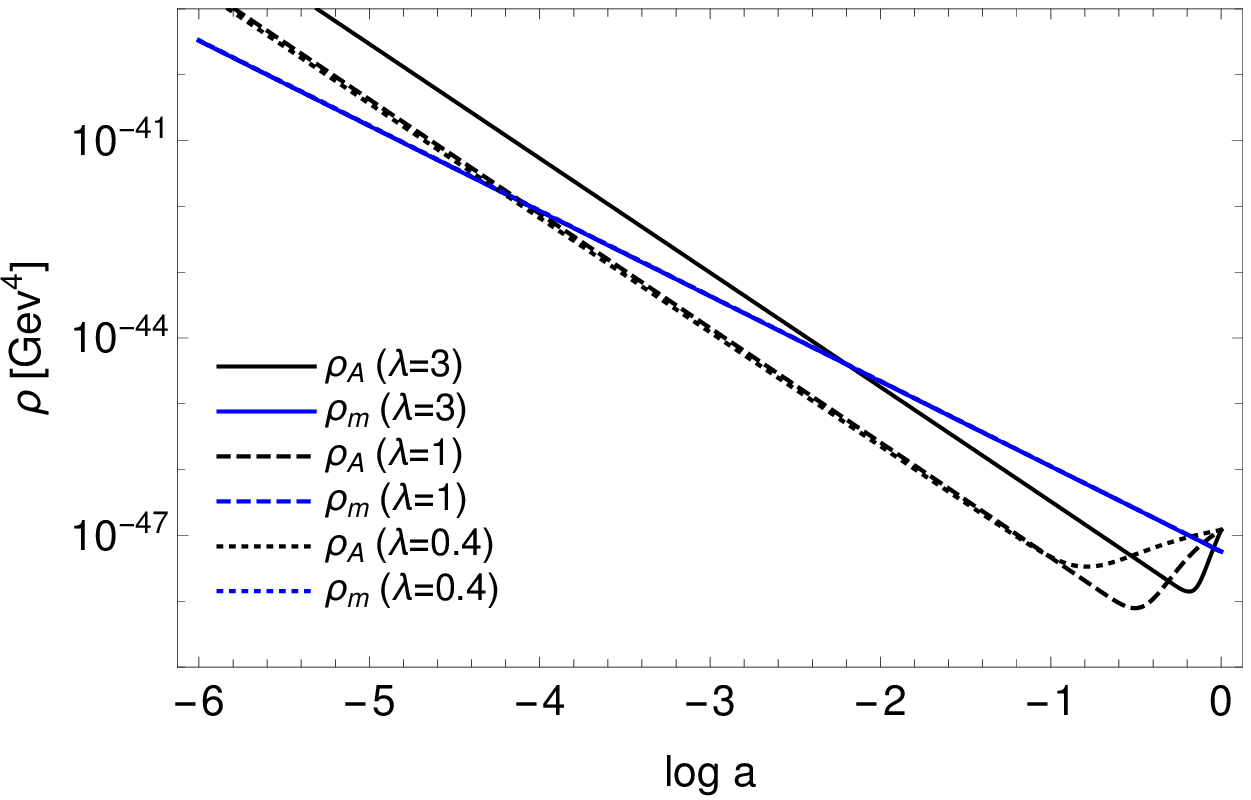}
\caption{Evolution of the state parameters versus the number of e-folds $N$ in the top and middle panels with $\lambda=1$ and 0.4, respectively, for different values of $q$ as can be read in the legends. Dashed and solid lines denote solutions for $w_{eff}$ and $w_{A}$ respectively. We show the evolution of the energy density for each component in the bottom panel for $q=-0.1$ and different values of $\lambda$ as indicated in the legends. We have chosen $u^{(0)}=0.9$ as a comparable value with respect to the $x^{(0)}$ and $y^{(0)}$ values.  The other initial conditions have been chosen to match the present value $\Omega_{A}=\Omega_{DE}^{(0)}=0.68$.}\label{fig:5}
\end{figure}
To see the dynamical evolution of the system characterized by such fixed points, we display the reduced phase space portrait of the dynamical system eqn.~(\ref{sec:4:eqn3}) in fig.~\ref{fig:2} for the same parameter values  as in fig.~\ref{fig:1} (except for $q$ that takes several values as indicated in the caption). These two-dimensional plots are obtained by setting the slice $u=u_{c}$ at the attractor point $D_{i}$.  Notice that a negative $u_{c}$ determines the physical region enclosing $D_{1}$ and $D_{2}$; on the contrary, we have the equivalent physical region enclosing the points $D_{3}$ and $D_{4}$ as can be ascertained in fig.~\ref{fig:3} (see also fig.~\ref{fig:1}). Namely, the corresponding phase spaces have the same physical meaning as we already discussed. For the sake of easiness, we restrict our attention to the case with negative $u$ in all the subsequent plots, which corresponds to portraying $D_{1,2}$ and the saddle points $A_\pm$ and $B_{-}$. Thus, all trajectories leaving the saddle point $A_{+}$ (dark radiation), with some of them passing close to $B_{-}$ (scaling solution) move towards one of the attractor solutions, $D_{1}$ or $D_{2}$, (downwards or upwards, respectively), which accounts for an accelerated expansion period as can be seen in fig.~\ref{fig:2}.  In the same plot, the light yellow contour indicates the region where acceleration is realizable which remarkably occurs for any set of values for the model parameters. In addition, we can see that the effect of varying $q$ is noticed only by a slight shift in $B_{-}$ and not on $D_{1,2}$.  It is evident that, for $q=0$, $B_{-}=B_{+}$ at $x_{c}=0$; this  uncoupled case corresponds to the top panel of fig.~\ref{fig:2}.

We end this part by analysing the effect of changing $\lambda$. As expected from Table~\ref{tab:table1}, the attractor points $D_{1,2}$ ($D_{3,4}$) move along the black boundary towards $A_-$ ($A_+$) when we increase $\lambda$. This behaviour can be appreciated in  fig.~\ref{fig:3} where we have taken $\lambda=0.8$ in contrast to the case $\lambda=0.4$ depicted in  fig.~\ref{fig:2}. As a further remark about this analysis, we note that each component of the total energy density is given by a well-characterized state parameter at each fixed point which means that there is no any dependence on the model parameters even though they control the whole dynamical behaviour of $w_{eff}$. The parameter $\lambda$ affects, for instance, the value of $w_{eff}$ at the onset of matter domination. These features will be appreciated in the next section when we numerically solve the cosmological evolution of the system in terms of the parameters of the model.

\section{Numerical results: Cosmological background evolution}\label{sec:4}
We employ numerical methods in this section to solve the cosmological evolution of the model in order to verify the dynamical behaviour found in the previous section. 
We start by exploring the viable region of the parameter space in accordance with the analysis already performed. 
After an exhaustive exploration of the numerical solutions in terms of the model parameters, we notice that the behaviour of the density parameters associated to the vector field $\Omega_{A}$ and the matter component $\Omega_{m}$ is mostly influenced at very high redshift (i.e., large and negative e-folding $N$) by the parameter $\lambda$ in the range $1/3<\lambda<1$, displaying an early matter-dark radiation equality (see middle and bottom panels of fig.~\ref{fig:4}). Meanwhile, for $\lambda>1$, they display important changes with respect to the former at very low redshift as can be observed in fig.~\ref{fig:4} top panel. Importantly, the most notorious difference among the numerical solutions shows at the full matter domination epoch: for $\lambda>1$ the matter density parameter is always below one because of the non-vanishing vector energy contribution leading, for instance, to $\Omega_{m}\sim0.9$ for the case $\lambda=3$ (see top panel of fig.~\ref{fig:4}). In contrast, for $\lambda<1$, 
$\Omega_{m} \approx 1$. In general, by increasing $\lambda$ an order of magnitude, the matter density is reduced roughly $10\%$. It is also worth pointing out that, for $\lambda>3$, the transition between matter and DE domination occurs at $z<0.1$, i.e. very close to the present epoch, resulting in a cosmic coincidence situation. In short, as $\lambda$ increases, the matter-dark energy equality is slightly shifted to the present epoch. Hence, there should be an upper limit for $\lambda$ to have a successful dynamic expansion such that the coincidence appears much less severe today. As a conservative limit we set $\lambda$ to be $\mathcal{O}(1)$ but a more stringent limit will be derived by the use of observational data. 

The effect of changing $q$ is notoriously amplified in the numerical solutions with $\lambda=1$ (compare the middle panel with the other two in fig.~\ref{fig:4}) and, in particular, around matter-dark radiation equality but with almost indistinguishable differences at the very early epoch and shortly before today even for any value of $\lambda$. A consequence of this is the convergence of the energy-density solutions for all $q$ values (including the uncoupled case $q=0$) at the aforementioned epochs as can be appreciated in the same figure. In other words, the effect of $q$ turns out to be noticeable in an intermediate stage of the universe evolution;  in contrast, it is turned off in other stages as we have already assessed in the dynamical system analysis presented in section \ref{sec:3}. Interestingly, all the numerical solutions exhibit, as a general trend, an early contribution of the vector field in the form of dark radiation and a late contribution in the form of DE. To put it another way, the kinetic energy and the potential of the vector fields dominate at early and late times respectively, with the possibility, provided that $q\neq0$, of having an early contribution of the cosmic triad to the total energy density during the matter domination.  This could leave some imprints on the structure formation that can be confronted with observations.
 
It is very instructive to display how the effective state parameter and the one for the vector field evolve with time.  This allows us to more clearly observe the dominance of each component, with particular attention on the tracking solutions $w_{r}\to w_{m}\to w_{DE}$ ensuring sequentially (dark) radiation, matter, and DE transitions. In fact, it is appealing to see how $w_{A}$ tracks $w_{eff}$ at very high and low redshifts. This result is shown in fig.~\ref{fig:5} for two particular values of $\lambda$. Both cases exhibit a well behaved transition of $w_{eff}$ from dark radiation to matter and then to DE domination. In particular, the case $\lambda=1$ in the top panel shows an absolute minimum for $w_{eff}$ before it reaches a value $w_{eff}=-1$.  The same panel shows, in addition, a deeper downfall of $w_{A}$ contrary to the case $\lambda=0.4$ in the middle panel. Indeed, the most significant effect of $\lambda$ on the dynamical evolution of the universe is towards the onset of the accelerated expansion epoch until the minimum of $w_{A}$ due to the dominance of the potential energy in such a period, as expected from the equations of state (\ref{sec:4:eqn5}) and (\ref{sec:4:eqn6}). In physical terms, the vector field displays a phantom behaviour before achieving the de Sitter regime. Later, after the oscillation, the vector field essentially freezes with $\rho_{A}=V=\rm const.$ and the Hubble parameter becomes constant as can be inferred from the density parameters shown in fig.~\ref{fig:4}. 

Once the phantom regime ($w_{A}<-1$) is achieved, one can think immediately of the arising of singularities and possible quantum and classical instabilities that may plague the dark energy model, features that are commonly manifested in phantom dark energy scenarios (see, for instance, ref. \cite{Nojiri:2005sx}). The phantom behaviour, however, does not always lead to instabilities as is the case, for instance, in coupled and multi-field dark energy models \cite{Sur:2008tc}. It is worth mentioning that the phantom barrier crossing can be achieved in a consistent way \cite{Caldwell:1999ew}, satisfying all the stability conditions of linear cosmological perturbations. In our case, the phantom stage is of no concern at all since, first of all, once the vector field crosses the barrier ($w_{A}=-1$), it quickly goes back to the phantom divide line\footnote{It directly implies that the Big Rip singularity is not present in this model which is a characteristic and an inherent finite-time singularity in phantom dark energy models \cite{Nojiri:2005sx} such as in k-essence theories.}, remaining forever there and, second, the period of time during which this phantom exists, before it fades away or may be exorcised, is very short compared to the late accelerated expansion period of the universe. We state then that the phantom regime exhibited in this model is not problematic both at the background and at the linear perturbative level.

To sum up things, the cosmic triad exposes two outstanding features during the evolution of the universe. At early times, during the radiation era, it tracks $w_{eff}\to w_{A}=1/3$, contributing thus to radiation in the form of dark radiation whereas, at late times, it changes its role to DE, providing a de Sitter-type accelerated solution with $w_{eff}\to w_{A}=-1$. Let us now point out that the effect of $q$ on the numerical solutions is rather significant for $w_{A}$ at the matter domination era in accordance with the dependence of the state parameter in eqn.~(\ref{sec:4:eqn6}) via the $q z^{2}$ term which accounts for the matter energy density. 

Finally, we plot the energy density for both components and investigate the effects of the model parameters during the evolution of the universe in the bottom panel of fig.~\ref{fig:5}. From here, we conclude that the only significant change occurs for $\rho_{A}$ (black curves); in particular the solution with $\lambda=3$ (solid black curve ) is quite distinguishable from the solutions $\lambda=0.4$ and $\lambda = 1$ along the whole evolution. These latter solutions are, however, distinct each other only at very low redshift as can be appreciated. Hence, comparing the distinct numerical solutions, we notice that those distinguishable features are due to the influence of the potential at low redshift, resulting in a deeper downfall for the cases $\lambda=1$ and $\lambda = 3$ consistent with the dynamics of $w_{A}$ (see, for instance, how deep is the solution with $\lambda=1$ in contrast to the one with $\lambda=0.4$). Interestingly, both $\rho_{A}$ and $\rho_{m}$ match twice: the first match (dark radiation-matter equality) occurs around $z\approx 60$ ($N=-\log(1+z)$) for dotted ($\lambda=0.4$) and dashed ($\lambda=1$) black curves and $z\approx 7$ for the solid ($\lambda=3$) black curve, whereas the second match (matter-dark energy equality) happens close to the present epoch as we have discussed when analysing the energy densities in fig.~\ref{fig:5}. We argue then that these solutions are not fined tuned in view of the coincidence problem due to the dominance of the potential during the accelerated period.

\section{Discussion and conclusions}\label{sec:5}


We proposed a novel coupling between multi-Proca vector fields and cold dark matter at the level of the action through a vector mass-type term. The general interacting term in a FLRW universe, with the cosmic triad configuration for the space-like vector fields, is of the form $Q=-\frac{3f_{,X}}{f}\rho_{m}$, it being fully independent of the vector field nature (i.e., the specific structure of the associated group is not manifest, 
see eqn.~(\ref{sec:2:eqn8})). 

The dynamical system analysis revealed the existence of three physical critical points corresponding to dark radiation behaviour with $\Omega_{A}=1$ (saddle point), coupled multi-Proca vector-matter behaviour or simply a kinetic scaling solution where the  density ratio scales as $\rho_{DE}/\rho_{m}\propto -2q$ (saddle point), and vector DE behaviour leading to an accelerated expansion period with $\Omega_{A}=\Omega_{DE}$ (de Sitter-type attractor). These transitions can easily be seen in the phase space orbits in figs.~\ref{fig:2} and \ref{fig:3}. 

For completeness, we also performed numerical computations of the density parameters and the effective state parameter to verify the evolution of the universe and to more clearly investigate the effects of the model parameters $q$ and $\lambda$. As an interesting result, we found an appealing behaviour for the energy density associated to the vector fields that can naturally alleviate the coincidence problem due to the emergence of the potential at low redshift with no fine tuning. 
Indeed, for $\lambda=3$, $\rho_{A}$ is at most two orders of magnitude larger than $\rho_{m}$ in the past (high redshift), those differences being even smaller across the whole cosmic history for $\lambda=1$ as can be seen in the bottom panel of fig.~\ref{fig:5}.

As future theoretical research, we propose building more general couplings involving, for instance, the field strength or its dual in order to make the structure of the group explicit. Moreover, 
though our proposal is not formally framed in the context of conformal transformations, it would be worth investigating the respective vector transformation in the context of vector-tensor theories as was recently done for U(1) vector fields \cite{Papadopoulos:2017xxx} and the extended vector-tensor theories \cite{Kimura:2016rzw}.

To conclude, we proposed in this work a novel coupling to matter that allows the multi-Proca vector fields to exist during the matter dominated epoch, exhibiting consequently a novel scaling solution which is cosmologically appealing in order to address the coincidence problem and the debated Hubble tension. Though we did not attack directly such an issue in this work, it would be worth examining in a future work whether this model may conciliate the observational data at high and low redshifts.  This would serve as a first challenge to put the model in contact with observational constraints. Therefore, the inclusion of radiation and baryonic components must be included for consistency in the cosmological parameter estimation from high precision cosmology. However,  as the latter are decoupled from the rest of the components, the main results obtained here, for the merely qualitative purposes, are quite general and the quantitative assessments are well estimated at low redshift within a small percent error.

On the other hand, the effects of the coupling could be important in the evolution of matter density perturbations and all the derived quantities associated with measurements of large-scale structures such as the growth rate and the redshift-space distortion $f\sigma_{8}$.
Hence, the subsequent observational test would be to constrain the evolution of matter perturbations and the evolution of gravitational potentials by including redshift-space distortion measurements from different observational surveys. Another important examination within the context of structure formation would be to investigate the influence of such a coupling on the spherical collapse and number counts at the corresponding redshift. Regarding the dark radiation effects in the early-universe physics (see e.g. refs. \cite{Ackerman:mha,Fischler:2010xz,Archidiacono:2011gq,Hasenkamp:2012ii,Baumann:2015rya}), they can imprint detectable signatures onto the cosmic microwave background radiation causing both spectral distortions and additional anisotropies.  In addition, they can contribute to the effective number of relativistic species $N_{eff}$ as was studied in the dark energy model of ref. \cite{Mehrabi:2017xga} whose action is composed of non-Abelian gauge fields. Our plan is to address these concerns in a forthcoming publication in order to guarantee the consistence of the model throughout the different stages of the universe. 
It will also be important to check whether the vector field coupling to  matter may induce some sort of  instability in the scalar sector that leads to additional constraints and to establish their consistency when contrasting with observational data.

\acknowledgments
L. G. G is supported by the  Postdoctoral Fellowship Programme N$^\circ$ $2020000102$ of the Vicerrector\'ia de Investigaci\'on y Extensi\'on - Universidad Industrial de Santander. This work was supported by the following grants: Patrimonio Autónomo - Fondo Nacional de Financiamiento para la Ciencia, la Tecnología y la Innovación Francisco José de Caldas (MINCIENCIAS - COLOMBIA) Grant No.  110685269447 RC-80740-465-2020, project 69553, and Dirección de Investigación y Extensión de la Facultad de Ciencias - Universidad Industrial de Santander Grant No. 2460. L. G. G acknowledges Departamento de F\'isica - Universidad del Valle for its hospitality at early stages of this project and in particular C\'esar A. Valenzuela-Toledo for stimulating discussions during his scientific visit.  L. G. G dedicates this work to the memory of his dear friend and colleague Diego Leonardo C\'aceres Uribe who passed away
so unexpectedly: your essence shall endure forever among us.

\appendix*

\section{Stability conditions}

We show in this appendix the general conditions that render the phase space portrait stable. The stability analysis follows the standard criteria of finding the eigenvalues of the Jacobian matrix $\mathcal{M}$ associated to the linear system and classifying them according to the stability criteria.  This allows us to establish the dynamical character of the fixed points and, therefore, their cosmological viability.

The matrix elements of $\mathcal{M}$ are:

\begin{figure*}[!htb]
\centering
\includegraphics[width=0.75
\hsize,clip]{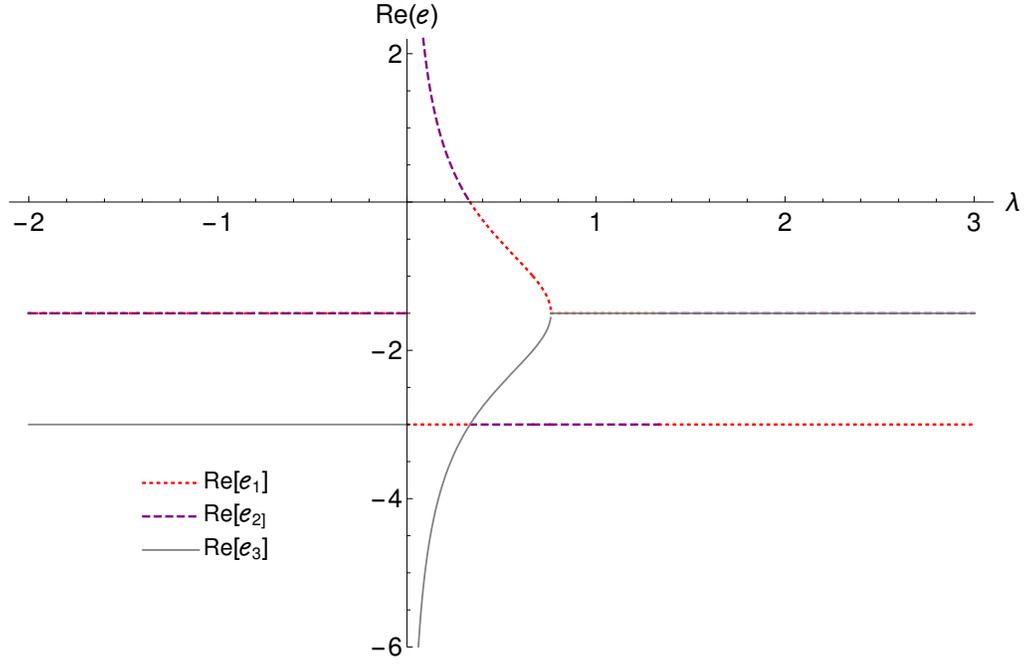}
\caption{Real part of the eigenvalues of the matrix $\mathcal{M}$ associated to the fixed points $D_{i}$ in one-dimensional phase space. Negative values for Re$[e_{i}]$ guarantee the stability of the critical points.}\label{fig:6}
\end{figure*}

\begin{align}
    \mathcal{M}_{11}&=\frac{2qx(\sqrt{2}-ux)}{u}-q(-1+x^{2}+y^{2})\nonumber\\ &+\frac{-1+3x^{2}-y^{2}(3+6u^{2} \lambda)}{2},\nonumber \\
    \mathcal{M}_{12}&=\frac{2 q (\sqrt{2}-ux)y}{u} + 
 6\sqrt{2}uy\lambda-xy(3+6u^{2}\lambda), \nonumber \\
 \mathcal{M}_{13}&=-\frac{qx(-1+x^{2}+y^{2})}{u}-\frac{
 q(\sqrt{2}-ux)(-1+x^{2}+y^{2})}{u^{2}}\nonumber\\
 &+3\sqrt{2}y^{2}\lambda-6uxy^{2}\lambda, \nonumber\\
 \mathcal{M}_{21}&=-y(-x+2qx+3\sqrt{2}u\lambda),\nonumber\\
 \mathcal{M}_{22}&=-y(2qy+3y(1+2u^{2} \lambda))\nonumber \\&+\frac{x^{2}-2q(-1+x^{2}+y^{2})}{2}-3\sqrt{2}ux\lambda \nonumber \\
 &-\frac{3(-1+y^{2})(1+2u^{2}\lambda)}{2},\nonumber\\
 \mathcal{M}_{23}&=-y(3\sqrt{2}x\lambda+6u (-1+y^{2})\lambda),\nonumber\\
  \mathcal{M}_{31}&=\sqrt{2},\nonumber\\
   \mathcal{M}_{32}&=0,\nonumber\\
      \mathcal{M}_{33}&=-1.\nonumber\\
 \end{align}
 
Thus, the fixed points and their characterization are as follows: 
\begin{itemize}
\item \textit{Kinetic dominated solution} ($A_{\pm}$): \\
$e_{1}=2\;, e_{2}=-1,\; e_{3}=1$.\\ These points clearly represent saddle points since the eigenvalues are real valued with two of them having opposite signs.
\item \textit{Kinetic scaling solution} ($B_{\pm}$):\\
$e_{1}=\frac{3}{2}, e_{2,3}=\frac{3-6q\pm\sqrt{7}\sqrt{-1 +4q-4q^{2}}}{4(-1+2q)}$.\\
It is straightforward to verify that the real parts of $e_{2,3}$ are always negative for any value of $q$, with the exception of $q\neq1/2$, such that these fixed points are also saddle points.
\item \textit{Vector dominated solution} ($D_{i}$):
The corresponding eigenvalues $e_{i}$ are
quite long and little illuminating to be reported here as a guidance for analytical treatment. Their real parts are plotted instead in terms of the parameter $\lambda$ in fig.~\ref{fig:6} where the conditions $\lambda<0$ and $\lambda>1/3$ are read off as a requirement for having ${\rm Re} [e_{i}]<0$. Hence, these fixed points are stable representing dynamical attractors. The condition $\lambda>1/3$ guarantees, in turn, their cosmological viability as can be seen in Table~\ref{tab:table1}.
\end{itemize}

\bibliography{apssamp}
\bibliographystyle{apsrev4-1}
\end{document}